\documentstyle[12pt,epsfig]{article}          

\def\gsim{\lower0.5ex\hbox{$\:\buildrel >\over\sim\:$}}
\def\lsim{\lower0.5ex\hbox{$\:\buildrel <\over\sim\:$}}

\def \rp{{R\hspace{-0.22cm}/}_P}
\def \lp{{L\!\!\!/}}
\def \anb{{\bar{\tilde\nu}}} 

\def \be {\begin{equation}}
\def \ee {\end{equation}}
\def \bea{\begin{eqnarray}}
\def \eea{\end{eqnarray}}

\def \n{\noindent}

\begin{document}
\baselineskip 18pt plus 2pt

\noindent \hspace*{10cm}UCRHEP-T235 (December 1998)\\
\noindent \hspace*{10cm}BNL-HET-98/42

\begin{center}
{\bf The flavor changing top decay
\boldmath{$t \to c \tilde\nu$} or \boldmath{$\tilde\nu \to t \bar c$} 
in the MSSM without \boldmath{$R$}--parity}
\vspace{.2in}

S. Bar-Shalom$^a$, G. Eilam$^b$ and A. Soni$^c$
\vspace{.2in}

$^a$ Physics Department, University of California, Riverside, 
CA 92521, USA\\
$^b$ Technion-Institute of Technology, Haifa 32000, Israel \\
$^c$ Physics Department, Brookhaven National Laboratory, Upton, NY 11973,
USA\\
\vspace{.4in}

{\bf Abstract}\\
\end{center}

Widths for the new flavor changing top quark decay 
$t \to c \tilde\nu$ or for the reversed sneutrino decay 
$\tilde\nu \to t \bar c$ are calculated
in the MSSM without $R$-parity conservation. 
For large $\tan\beta$, {\it e.g.,}   
$\tan\beta \sim m_t/m_b \sim 40$,  
${\rm Br}(t \to c \tilde\nu) > 10^{-5}$ or 
${\rm Br}(\tilde\nu \to t \bar c) > 10^{-6}$ in a relatively 
wide range of the supersymmetric parameter space as long as there is 
more than one non-zero $R$-parity violating coupling.
In the best cases, 
with a typical squark mass around 100 GeV, we find that 
${\rm Br}(t \to c \tilde\nu) \sim 10^{-4} - 10^{-3}$ or 
${\rm Br}(\tilde\nu \to t \bar c) \sim 10^{-5} - 10^{-4}$.
For $\tan\beta \sim {\cal O}(1)$ the corresponding branching 
ratios for both top or sneutrino decays 
are too small to be measured at the LHC. Therefore, the decays 
$t \to c \tilde\nu$ or $\tilde\nu \to t \bar c$ 
appear to be sensitive to $\tan\beta$ and may be detected at 
the LHC. The branching ratios of the corresponding decays to an up 
quark instead of a charm quark, {\it e.g.,}     
$t \to u \tilde\nu$ or 
$\tilde\nu \to t \bar u$, may also be similar.       

\newpage

\n \underline{\bf 1. Introduction} \\

Flavor changing decays of heavy quarks which have been the subject
of intense theoretical activity for a long time,
are especially significant 
as they provide important tests of the Standard Model (SM).
The extraordinary large mass of the top renders the GIM mechanism 
very effective in the SM, so that the flavor changing top decays are highly 
suppressed \cite{prd44p1473}. 
Experimental searches for the flavor changing decays of 
the top are therefore very good probes of new physics. 
It is expected that the LHC will produce about $10^7 -10^8$ $t \bar t$ 
pairs, therefore rare top decays with branching ratios around $10^{-6}$ 
should be accessible at the LHC. 

In the SM, the decays $t \to cV~(V=\gamma,~Z,~{\rm or}~g)$
and $t \to c H^0$, have branching ratios (Br)
much smaller than $10^{-6}$ 
\cite{prd44p1473,fcsm} and should therefore be very useful in 
searching for new physics at the LHC. 
In some extensions of the SM such as multi-Higgs doublets models 
\cite{prd44p1473,plb307p387} and the MSSM \cite{tcvmssm},  
${\rm Br}(t \to cV)$ can be several orders of magnitude larger than 
their SM value.      
However, although in the MSSM ${\rm Br}(t \to cg) \sim 10^{-5}$ may be 
possible, for $t \to c \gamma$ and $t \to c Z$, typically, 
${\rm Br}(t \to c \gamma,~cZ) \lsim 10^{-6}$ 
\cite{prd44p1473,plb307p387,tcvmssm}.

The most optimistic results for
 ${\rm Br}(t \to c V)$ were reported in \cite{tcvrpv}, 
where it was shown that, in the MSSM without 
$R$-parity, these branching ratios can be as large as 
${\rm Br}(t \to c g) \sim 10^{-3}$, 
${\rm Br}(t \to c Z) \sim 10^{-4}$ and 
${\rm Br}(t \to c \gamma) \sim 10^{-5}$ if 
the squarks have masses not much larger than 100 GeV.

For the decay $t \to cH^0$ the theoretical predictions for the 
branching ratio span several orders of magnitude as one considers 
beyond the SM scenarios. While in the MSSM 
${\rm Br}(t \to c H^0)$ can reach $\sim 10^{-5}$ at the most 
\cite{prd49p3412},
in a class of multi-Higgs doublets models in which the neutral Higgs 
can have a non-vanishing tree-level coupling to a $t \bar c$ pair, 
${\rm Br}(t \to c H^0)$ around $10^{-2}$ is not ruled out \cite{2hdm3}.    

In this paper we explore the new flavor changing 
top decays $t \to c \tilde\nu,~t \to c \anb$ 
($\tilde\nu$=sneutrino, $\anb$=anti-sneutrino).
These decays require lepton number to be violated and, therefore,
cannot occur within supersymmetry in its minimal version. 
We, therefore, consider $t \to c \tilde\nu,~c \anb$ or 
$\tilde\nu,~\anb \to t \bar c,~\bar t c$ (depending on whether 
$m_t > m_{\tilde\nu}$ or $m_t < m_{\tilde\nu}$, respectively) in the 
MSSM with $R$-parity violation ($\rp$) \cite{rpreview}. 
 
The $\rp$ interaction of interest to us is 
the $\lambda^{\prime}$ type --    
lepton number violating -- operator \cite{rpreview}:

\be
W_{\lp} = \lambda_{ijk}^{\prime} {\hat L}_i  
{\hat Q}_j {\hat D}_k^c \label{rplag1}~,     
\ee  

\n and the corresponding trilinear soft breaking operator:

\be
{\cal L}_{\lp}^{soft} = \lambda_{ijk}^{\prime} A^{\prime} {\tilde \ell}_i   
{\tilde q}_j {\tilde d}_k^c \label{rplag2}~,     
\ee

\n where in Eq.~\ref{rplag1} 
${\hat L}$ and ${\hat Q}$ are the SU(2)--doublet lepton and quark 
superfields, respectively and ${\hat D}^c$ are the quark singlet 
superfields. In Eq.~\ref{rplag2} ~$\tilde\ell$, $\tilde q$ and $\tilde d$ 
are the sleptons, left handed squarks and right handed down-squarks 
corresponding to the superfields ${\hat L}$, ${\hat Q}$ and ${\hat D}$, 
respectively. Also, for simplicity, the 
trilinear soft breaking couplings in Eq.~\ref{rplag2} have been defined 
to include the Yukawa type $\lambda^{\prime}$ couplings. Notice that  
$A^{\prime}$ has a mass dimension and should naturally attain values 
of the order of the typical squark mass or equivalently the typical 
supersymmetric (SUSY) 
mass scale (see {\it e.g.,} \cite{abel} and references therein).          

It is important to note that, on a purely phenomenological level, 
since there is no good theoretical reason which forbids $\rp$ SUSY models, 
the $\lambda^{\prime}$ couplings in Eqs.~\ref{rplag1} and 
\ref{rplag2} are a-priori expected to be of ${\cal O}(1)$
(see \cite{foot1}).
Existing bounds on the $\lambda^{\prime}$ couplings from 
low energy processes suggest that, typically, $\lambda^{\prime}_{ijk} < 0.5$ 
for squark and slepton masses of 100 GeV.
We wish to emphasize that these bounds are usually based on the 
assumption that an $\rp$ coupling, {\it e.g.,} 
$\lambda^{\prime}_{ijk}$ is non-zero only for one combination of 
the indices $ijk$ \cite{rpreview}. 
Indeed, if this assumption is relaxed, then in most cases             
 $\lambda^{\prime}_{ijk} \sim {\cal O}(1)$ cannot be ruled out. 

Another useful observation is that in the presence of $\rp$ interactions, 
existing limits on the SUSY spectrum which are obtained from 
high energy collider experiments become less restrictive, since in this model 
the LSP is no longer stable. That is, the distinct missing energy signal 
associated with the LSP in the MSSM with $R$-parity conservation is lost. 
Thus, for example, squark masses of the order of $\sim 100$ GeV 
are still allowed by present data \cite{rpreview}. 
           
With that in mind, in this work we show that if $m_t > m_{\tilde\nu}$, then 
${\rm Br}(t \to c \tilde\nu) > 10^{-5}$ is attainable 
within a relatively wide range 
of the relevant SUSY parameter space and may reach 
$10^{-4} - 10^{-3}$ for $\tan\beta \sim m_t/m_b \sim 40$ 
(recall that $\tan\beta$ is the ratio between the vacuum expectation 
values of the two Higgs fields in the model) and with 
the typical squark mass around 100 GeV.
Similarly, in the case $m_t < m_{\tilde\nu}$, we find that 
${\rm Br}(\tilde\nu \to t \bar c) > 10^{-6}$ is possible and can reach 
$10^{-5} - 10^{-4}$ if, again, $M_s \sim 100$ GeV (where $M_s$ denotes
the squark mass) and $\tan\beta \sim m_t/m_b \sim 40$. 
In fact, if indeed some of the $\lambda^{\prime}$ couplings are saturated to 
be of order one, then the LHC can effectively serve as a sneutrino factory 
having about $10^8-10^9$ sneutrinos with 
a mass of $200 - 300$ GeV produced inclusively \cite{sneutogg}. 
Therefore, if $m_t < m_{\tilde\nu}$, the above 
flavor changing sneutrino decay, $\tilde\nu \to t \bar c$, 
appears to be as useful 
as the top decay $t \to c \tilde\nu$, which is of course 
relevant only for the corresponding mass range $m_t > m_{\tilde\nu}$. 
          
We also note that, for an up quark instead of a charm quark,
the branching ratios remain practically the same, {\it i.e.,} 
${\rm Br}(t \to u \tilde\nu) \simeq {\rm Br}(t \to c \tilde\nu)$ or   
${\rm Br}(\tilde\nu \to t \bar u) \simeq {\rm Br}(\tilde\nu \to t \bar c)$ 
if the $\rp$ couplings relevant for the two cases are the same. 
 
The paper is organized as follows: in section 2 we present 
an analytical derivation and the numerical results  
for the branching ratio of $t \to c \tilde\nu,~c \anb$
in the MSSM with $\rp$. 
In section 3 we evaluate the sneutrino and anti-sneutrino decays 
$\tilde\nu,~\anb \to t \bar c,~\bar t c$ in the case 
$m_t < m_{\tilde\nu}$ and in section 4 we summarize. \\  

\n \underline{\bf 2. The case \boldmath{$m_t > m_{\tilde\nu}$} and 
the top decays \boldmath{$t \to c \tilde\nu$},  
\boldmath{$t \to c \anb$}} \\

\begin{figure}[htb]
\psfull
 \begin{center}
  \leavevmode
  \epsfig{file=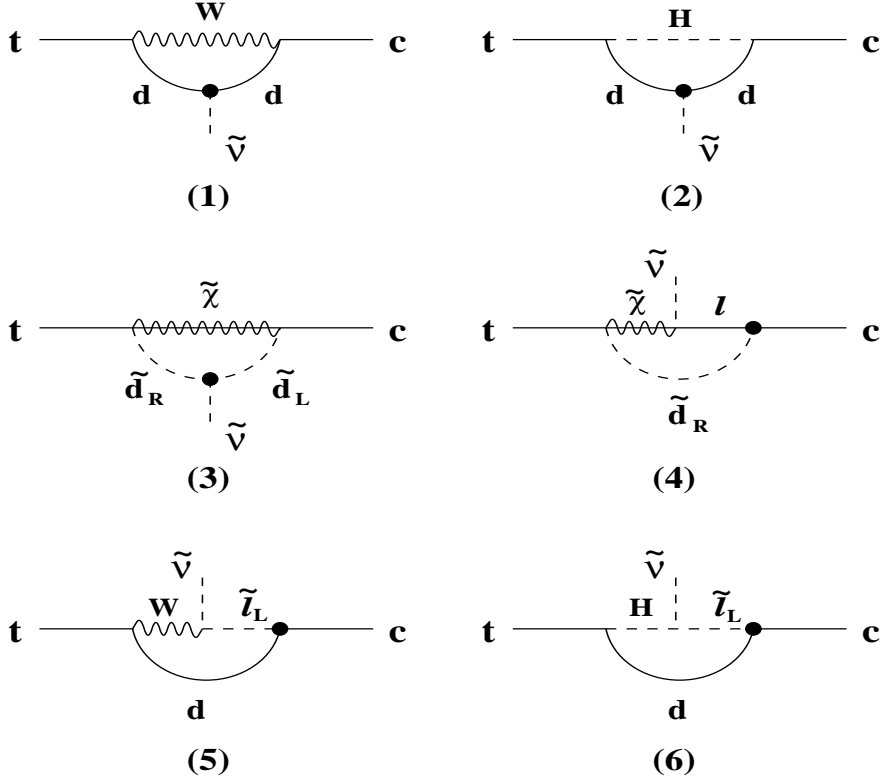,height=9.5cm,width=10cm,bbllx=0cm,bblly=2cm,bburx=20cm,bbury=25cm,angle=0}
 \end{center}
\caption{\emph{The one-loop Feynman diagrams for the decay 
$t \to c \tilde\nu$, induced by an insertion of one lepton number violating 
$\lambda^{\prime}$ coupling. 
The heavy dot denotes the lepton number violating vertex. 
See also \cite{footfig1}}.}  
\label{figure1}
\end{figure}

The relevant Feynman diagrams responsible for the decay $t \to c \tilde\nu$ 
at the one-loop level and with insertion of only one $\rp$ coupling 
are depicted
in Fig.~\ref{figure1}. 
Those penguin-like diagrams give  
rise to an effective amplitude for the decay $t \to c \tilde\nu$ 
of the form:
 
\be
-i {\cal M}^{\tilde\nu} = {\cal C}^{\tilde\nu} \bar u_c \left[ 
\left( \frac{m_c}{m_W} \right) \sum_\alpha {\cal A}_\alpha^{\tilde\nu} L +
\left( \frac{m_t}{m_W} \right) \sum_\alpha {\cal B}_\alpha^{\tilde\nu} R \right] u_t 
 \label{ttocs} ~,
\ee

\n where $L(R)=(1-(+) \gamma_5)/2$ and: 

\be
{\cal C}^{\tilde\nu} \equiv \frac{\alpha}{8 \pi s_W^2} \frac{m_{d_k}}{m_W} 
\lambda^{\prime \ast}_{ijk} V_{tk} V^{\ast}_{cj} \label{csneu}~.
\ee  
 
\n Also, here $s_W$ is the $\sin$e of the weak mixing angle and 
$m_{d_k}$ is the mass of the down quark of the $k$th generation ($k=1,2$ or 3).
The form factors $A_\alpha^{\tilde\nu}$ and $B_\alpha^{\tilde\nu}$
corresponding to diagrams $\alpha=1-6$ in Fig.~\ref{figure1} are given by 
(keeping all the masses):

\bea
&& {\cal A}_1^{\tilde\nu}= m_t^2 (C_{21}^1 +C_{22}^1 -2C_{23}^1 ) -
2m_W^2 (C_{0}^1 +C_{12}^1) - m_c^2 C_{22}^1 +2 C_{24}^1 - {\tilde C}_{12}^1
\label{starta}~,\\
&& {\cal A}_2^{\tilde\nu}= m_t^2 (1+ \cot ^2 \beta) (C_{12}^2 -C_{11}^2) 
- m_c^2 C_{12}^2 - m_{d_k}^2 \tan ^2 \beta C_{12}^2 - {\tilde C}_{0}^2
~,\\
&& {\cal A}_3^{\tilde\nu}= -A^{\prime} \sum_{m=1,2} ~\frac{Z_{2m}^-}{\cos\beta} \left( 
\frac{Z_{2m}^+}{\sin\beta} m_{{\tilde\chi}_m} C_{0}^3 + \sqrt 2  
Z_{1m}^{- \ast} m_W C_{12}^3 \right) 
~,\\
&& {\cal A}_4^{\tilde\nu}= \sum_{m=1,2} ~ \frac{Z_{2m}^-}{\cos\beta} \left( 
m_{\ell_i}^2 \frac{Z_{2m}^{- \ast}}{\cos\beta} C_{12}^4 - 
\sqrt 2  Z_{1m}^{+} m_W m_{{\tilde\chi}_m} (C_0^4 + C_{12}^4) \right)
~,\\  
&& {\cal A}_5^{\tilde\nu}= (m_t^2 -m_c^2)(C_{22}^5 - C_{23}^5) - 
m_{\tilde\nu_i}^2 (C_{22}^5 + C_{23}^5) +m_W^2 (C_{12}^5 + 2 C_{0}^5) 
\nonumber \\
&& ~~~~~~~~ -2C_{24}^5 -{\tilde C}_{12}^5
~,\\
&& {\cal A}_6^{\tilde\nu}= (m_W^2 \sin 2 \beta - m_{\ell_i}^2 \tan\beta) 
\tan\beta C_{12}^6 \label{enda}~,
\eea  

\n and:

\bea
&& {\cal B}_1^{\tilde\nu}= (m_t^2 - m_{\tilde\nu_i}^2) 
( 2C_{23}^1 -C_{21}^1 -C_{22}^1) -
2m_W^2 (C_{11}^1 -C_{12}^1) + m_c^2 (C_{22}^1-C_{21}^1) \nonumber\\  
&&~~~~~~~~  +{\tilde C}_{12}^1 - {\tilde C}_{11}^1
\label{startb}~,\\
&& {\cal B}_2^{\tilde\nu}= m_{d_k}^2 \tan^2 \beta 
(C_{12}^2 - C_{11}^2 -C_{0}^2) - m_{d_k}^2 C_0^2 - m_c^2 (C_0^2 + C_{11}^2) 
\nonumber\\
&&~~~~~~~~ - m_c^2 \cot^2 \beta (C_0^2 + C_{12}^2)
~,\\
&& {\cal B}_3^{\tilde\nu}= \sqrt 2 A^{\prime} m_W \sum_{m=1,2} ~
\frac{Z_{2m}^- Z_{1m}^-}{\cos\beta} (C_{12}^3 - C_{11}^3) 
~,\\
&& {\cal B}_4^{\tilde\nu}= \sum_{m=1,2} ~\frac{Z_{2m}^-}{\cos\beta} \left( 
\sqrt 2  Z_{1m}^{+} m_W m_{{\tilde\chi}_m} (C_0^4 + C_{11}^4)
- m_{\ell_i}^2 \frac{Z_{2m}^{- \ast}}{\cos\beta} C_{11}^4  \right)
~,\\  
&& {\cal B}_5^{\tilde\nu}= (m_t^2 -m_c^2)(C_{21}^5 - C_{23}^5) + 
m_{\tilde\nu_i}^2 (C_{21}^5 + C_{23}^5) - m_W^2 (C_{11}^5 + 2 C_{0}^5)
\nonumber \\ 
&&~~~~~~~~+2C_{24}^5 +{\tilde C}_{11}^5
~,\\
&& {\cal B}_6^{\tilde\nu}= - (m_W^2 \sin 2 \beta - m_{\ell_i}^2 \tan\beta) 
\left ( \tan\beta (C_{11}^6 +C_0^6) + \cot\beta C_0^6 \right) 
\label{endb} ~, 
\eea     

\n where the lepton flavor ($\ell_i$) as well as the sneutrino flavor 
($\tilde\nu_i$), $i=1,2$ or 3, are set by the choice 
of the first index in the $\rp$ coupling $\lambda^{\prime}_{ijk}$.
Also, $m_{{\tilde\chi}_m}$, $m=1,2$, are the chargino masses and 
$Z^\pm$ are the $2 \times 2$ matrices that diagonalize the charginos 
$2 \times 2$ mass matrix (see {\it e.g.}, Ref. \cite{ourtbpaper}).
The $C_0^\alpha$, ${\tilde C}_0^\alpha$, $C_{pq}^\alpha$ 
and ${\tilde C}_{pq}^\alpha$, $p=1,2$ and $q=1-4$,
 in Eqs.~\ref{starta} - \ref{endb} are 
the three-point loop form factors associated with the loop integrals of 
diagrams $\alpha= 1-6$ in Fig.~\ref{figure1}. Using the notation in \cite{ourtbpaper}, 
they are given by:

\bea
&& C^1 \equiv 
C^1 \left(m_W^2,m_{d_j}^2,m_{d_k}^2,m_t^2,m_{\tilde\nu_i}^2,m_c^2 \right)~,\\
&& C^2 \equiv 
C^2 \left(m_{H^+}^2,m_{d_j}^2,m_{d_k}^2,m_t^2,m_{\tilde\nu_i}^2,
m_c^2 \right)~,\\
&& C^3 \equiv 
C^3 \left(m_{{\tilde\chi}_m}^2,m_{{\tilde d}_{Rj}}^2,m_{{\tilde d}_{Lk}}^2,
m_t^2,m_{\tilde\nu_i}^2,m_c^2 \right)~,\\
&& C^4 \equiv 
C^4 \left(m_{{\tilde\chi}_m}^2,m_{{\tilde d}_{Rj}}^2,m_{\ell_i}^2,
m_t^2,m_c^2,m_{\tilde\nu_i}^2 \right)~,\\
&& C^5 \equiv 
C^5 \left(m_W^2,m_{d_j}^2,m_{{\tilde \ell}_{Li}}^2,m_t^2,m_c^2,
m_{\tilde\nu_i}^2 \right)~,\\
&& C^6 \equiv 
C^6 \left(m_{H^+}^2,m_{d_j}^2,m_{{\tilde \ell}_{Li}}^2,m_t^2,m_c^2,
m_{\tilde\nu_i}^2 \right)~.
\eea

\n Note that since the arguments are the same for all the $C^\alpha$'s 
belonging to diagram $\alpha$, 
we have omitted their subscripts and tildes.
$m_{{\tilde d}_{Rj}}$ and $m_{{\tilde d}_{Lk}}$ are the masses 
of the right handed and left handed squarks of the $j$ and $k$ generation 
(set by the second and third index in $\lambda^{\prime}$), 
respectively. Also, $m_{{\tilde \ell}_{Li}}$ is the mass of the left handed 
slepton of flavor $i$ corresponding to the first index in $\lambda^{\prime}$, 
and $m_{H^+}$ is the charged Higgs boson mass.  
  
Notice that ${\tilde C}_0,~{\tilde C}_{11},~{\tilde C}_{12}$ and 
$C_{24}$ contain ultraviolet divergences which, in dimensional 
regularization, appear as $1/\epsilon$ where $\epsilon = 4-D$ 
in $D$ dimensions. Of course, $\sum_{\alpha=1}^6 {\cal A}_\alpha$ and 
$\sum_{\alpha=1}^6 {\cal B}_\alpha$ should be free of those infinities; 
indeed, as can be checked analytically from Eqs.~\ref{starta}-\ref{endb},
those infinities cancel out.

The width for $t \to c \tilde\nu_i$ is then given by:

\bea
&& \Gamma(t \to c \tilde\nu_i) = \frac{ |{\cal C}^{\tilde\nu}|^2}{32 \pi} 
\frac{\omega(m_t^2,m_c^2,m_{{\tilde\nu}_i}^2)}{m_t^3} \frac{m_t^2m_c^2}{m_W^4} 
\times \nonumber \\
&&~~~~~~~~ \left\{ \left( m_t^2 + m_c^2 - m_{{\tilde\nu}_i}^2 \right) 
\left( \frac{m_W^2}{m_t^2} |\bar {\cal A}^{\tilde\nu}|^2 +
\frac{m_W^2}{m_c^2} |\bar {\cal B}^{\tilde\nu}|^2 \right)
+4 m_W^2 \Re{\rm e} \left[\bar {\cal A}^{\tilde\nu} 
(\bar {\cal B}^{\tilde\nu})^\ast \right] 
\right\} \label{sneuwidth}~,
\eea 

\n where we have defined:

\be
\bar {\cal A}^{\tilde\nu} \equiv \sum_{\alpha=1}^6 
{\cal A}_\alpha^{\tilde\nu} ~~,~~ 
\bar {\cal B}^{\tilde\nu} \equiv \sum_{\alpha=1}^6 
{\cal B}_\alpha^{\tilde\nu} \label{defAB1}~,
\ee

\n and:

\be
\omega(a^2,b^2,c^2) \equiv \sqrt { \left( a^2 - (b+c)^2 \right) 
\left( a^2 - (b-c)^2 \right) } \label{omega}~.
\ee 

For the top decay to an anti-sneutrino, $t \to c \anb$, 
the amplitude can be similarly written as:

\be
-i {\cal M}^{\anb} = {\cal C}^{\anb} \bar u_c \left[ 
\left( \frac{m_c}{m_W} \right) \sum_\alpha {\cal A}_\alpha^{\anb} 
L +
\left( \frac{m_t}{m_W} \right) \sum_\alpha {\cal B}_\alpha^{\anb} 
R \right] u_t 
 \label{ttocsstar} ~,
\ee 

\n where now:

\be
{\cal C}^{\anb} \equiv 
\frac{\alpha}{8 \pi s_W^2} \frac{m_{d_k}}{m_W} 
\lambda^{\prime}_{ijk} V_{tj} V^{\ast}_{ck} \label{csneustar}~,
\ee

\n and the form factors $A_\alpha^{\anb}$, 
$B_\alpha^{\anb}$, for $\alpha=1-6$, are related to 
$A_\alpha^{\tilde\nu}$, $B_\alpha^{\tilde\nu}$ 
as follows:

\bea
&&A_\alpha^{\anb} = \left[ B_\alpha^{\tilde\nu}  
\left( m_t \leftrightarrow m_c \right) \right]^\ast \label{rel1}~,\\
&&B_\alpha^{\anb} = \left[ A_\alpha^{\tilde\nu} 
\left( m_t \leftrightarrow m_c \right) \right]^\ast \label{rel2}~.
\eea 

\n The width $\Gamma(t \to c \anb_i)$ is then given 
by Eq.~\ref{sneuwidth} with the replacements  
${\cal C}^{\tilde\nu} \longrightarrow {\cal C}^{\anb}$,
$\bar {\cal A}^{\tilde\nu} \longrightarrow \bar {\cal A}^{\anb}$
and 
$\bar {\cal B}^{\tilde\nu} \longrightarrow \bar {\cal B}^{\anb}$, 
where:

\be
\bar {\cal A}^{\anb} \equiv \sum_{\alpha=1}^6 
{\cal A}_\alpha^{\anb}~~,~~ 
\bar {\cal B}^{\anb} \equiv \sum_{\alpha=1}^6 
{\cal B}_\alpha^{\anb} \label{defAB2}~,
\ee

\n and ${\cal A}_\alpha^{\anb}$ and 
${\cal B}_\alpha^{\anb}$ 
are given in Eqs.~\ref{rel1} and \ref{rel2}, respectively. 

We are now interested in calculating the branching ratio for a 
top quark to decay to $c \tilde\nu_i$ and to $c \anb$. 
In what follows we will assume that $m_{H^+} > m_t$, therefore, the branching 
ratio in question is given by: 

\be
{\rm Br}_i \equiv \frac{\Gamma(t \to c \tilde\nu_i) + 
\Gamma(t \to c \anb_i)}
{\Gamma(t \to b W^+)} \label{br}~.       
\ee 

\n It is clear from Eq.~\ref{csneu} that  
$\Gamma(t \to c \tilde\nu_i)$ is largest for $j=2$ and $k=3$, {\it i.e.,} for 
$\lambda^{\prime}_{i23}$, 
in which case it is proportional to the diagonal CKM elements 
$V_{tb}V_{cs}^{\ast} \sim 1$. Similarly, from Eq.~\ref{csneustar} we see that
$\Gamma(t \to c \anb_i)$ is largest for $j=3$ and $k=2$, being  
suppressed by the small off-diagonal CKM elements 
if $\lambda^{\prime}_{ijk} \neq \lambda^{\prime}_{i32}$.
Therefore, taking $\lambda^{\prime}_{i23} \neq 0$ for $i=1,2$ or 3, 
one finds that  
$\Gamma(t \to c \tilde\nu_i) > > \Gamma(t \to c \anb_i)$, 
leading to
${\rm Br}_i \approx \Gamma(t \to c \tilde\nu_i)/\Gamma(t \to b W^+)$.
In fact, for a given sneutrino flavor $i$, even if one allows 
$|\lambda^{\prime}_{i23}|=|\lambda^{\prime}_{i32}| \neq 0$, 
in which case the leading contribution to 
both $\Gamma(t \to c \tilde\nu_i)$ and 
$\Gamma(t \to c \anb_i)$ is proportional to 
$V_{tb}V_{cs}^{\ast}$, one still finds that $\Gamma(t \to c \tilde\nu_i)/  
\Gamma(t \to c \anb_i) \approx 
\left( {\cal C}^{\tilde\nu} / {\cal C}^{\anb} \right)^2 
\simeq m_b^2/m_s^2 > > 1$.

In the one coupling scheme, 
{\it i.e.,} only one $\lambda^{\prime}$ is non-zero, 
and if the relative mixing in the quark sector is solely due to absolute
mixing in the up-quark sector, the existing $1 \sigma$ bounds are
$\lambda^{\prime}_{i23} \lsim 0.2$ for sleptons and squarks masses of 
100 GeV \cite{rpreview}. These bounds are obtained from data on 
$D^0 - {\bar D}^0$ mixing \cite{agashe} and $D$ decays \cite{ref36dreiner}. 
However, if the one coupling scheme assumption is relaxed, then these bounds 
no longer apply due to possible cancelations between the  
$\lambda^{\prime}$'s with different indices (for details see \cite{agashe}). 

In fact, the branching ratio stays practically the same if, 
for example, some other   
$\lambda^{\prime}_{ijk}$ with $j \neq 2$ and $k \neq 3$ are also non-zero, 
since their contribution to $\Gamma(t \to c \tilde\nu_i)$ is suppressed
by off diagonal CKM elements.
Therefore, for our purpose it is sufficient to assume that there are only 
three non-zero couplings: $\lambda^{\prime}_{i23} \neq 0$ for $i=1,2,3$. 
The decays to an $e$-sneutrino, $\mu$-sneutrino and to a
$\tau$-sneutrino occur with the same rate if the three sneutrinos 
are degenerate with a mass $< m_t$ and if 
$|\lambda^{\prime}_{123}|=|\lambda^{\prime}_{223}|=|\lambda^{\prime}_{323}|$. 
In that case, ${\rm Br} \equiv {\rm Br}_1={\rm Br}_2={\rm Br}_3$ and 
the branching ratio for a top quark to decay to all sneutrino 
flavors is simply the sum $\sum_{i=1}^3 {\rm Br}_i = 3 {\rm Br}$.
For definiteness, and without loss of generality, 
throughout our analysis we present results for a given sneutrino flavor, 
say the $\tau$-sneutrino ($i=3$), and we drop the sneutrino index $i$.
We take $|\lambda^{\prime}_{323}| \equiv \lambda^{\prime}=1$ 
(which, as mentioned above, 
is not ruled out once the one coupling scheme approach is not realized), 
thus scaling out   
the $\rp$ coupling from the branching ratio and presenting results 
for: 

\be   
B^t \equiv {\rm Br}/ |\lambda^{\prime}|^2
\label{bt}~,
\ee

\n where Br is defined in Eq.~\ref{br}.

\begin{figure}[htb]
\psfull
 \begin{center}
  \leavevmode
  \epsfig{file=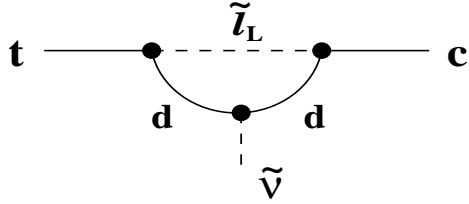,height=5cm,width=5cm,bbllx=0cm,bblly=2cm,bburx=20cm,bbury=25cm,angle=0}
 \end{center}
\caption{\emph{The one-loop Feynman diagram for the decay 
$t \to c \tilde\nu$, induced by an insertion of three lepton number violating 
$\lambda^{\prime}$ couplings. 
The heavy dots denote the lepton number violating vertices.}}  
\label{figure2}
\end{figure}

Before we continue we note that, to one-loop order, 
there is one additional diagram contributing to $t \to c \tilde\nu$ 
and similarly to $t \to c \anb$. 
This diagram is shown in Fig.~\ref{figure2} and 
is purely $\rp$ since it involves insertion of three $\rp$ couplings.
The amplitude corresponding to this diagram has the same structure as in 
Eq.~\ref{ttocs}, with the replacement:

\be 
{\cal C}^{\tilde\nu} \longrightarrow - 
\frac{\lambda^{\prime \ast}_{\ell n j} \lambda^{\prime}_{\ell p k}}{16 \pi^2}
\lambda^{\prime \ast}_{ijk} \frac{m_{d_j}}{m_W} V_{tp} V_{cn}^\ast 
\label{c3rp}~,
\ee

\n and with the form factors:

\bea
&& {\cal A}^{\tilde\nu}= m_W^2 C_{12}~,\\
&& {\cal B}^{\tilde\nu}= m_W^2 (C_0 +C_{11} - C_{12})~,
\eea

\n where:

\be
C \equiv C 
\left(m_{{\tilde \ell}_\ell}^2,m_{d_k}^2,m_{d_j}^2,m_t^2,
m_{\tilde\nu_i}^2,m_c^2 \right)~.
\ee

\n For $t \to c \anb$ we get:

\be 
{\cal C}^{\anb} \longrightarrow - 
\frac{\lambda^{\prime \ast}_{\ell n k} \lambda^{\prime}_{\ell p j}}{16 \pi^2}
\lambda^{\prime}_{ijk} \frac{m_{d_j}}{m_W} V_{tp} V_{cn}^\ast \label{cs3rp}~,
\ee

\n and, again, the corresponding form factors ${\cal A}^{\anb}$ and 
 ${\cal B}^{\anb}$ are extracted from 
${\cal A}^{\tilde\nu}$ and 
 ${\cal B}^{\tilde\nu}$ using the relations in  
Eqs.~\ref{rel1} and \ref{rel2}.

From Eqs.~\ref{c3rp} and \ref{cs3rp} we see that the amplitude of the 
diagram in Fig.~\ref{figure2} is largest when $j=3$ ($m_{d_j}=m_b$) and 
for $p=3$, $n=2$ ($V_{tp}V_{cn}^{\ast}=V_{tb}V_{cs}^{\ast}$). 
This will require 
both couplings of the type $\lambda^{\prime}_{\ell 2k}$ and 
$\lambda^{\prime}_{ \ell 3 k}$ to be non-zero. 
However, even within such a coupling scenario,  
we find that the contribution of the pure $\rp$ diagram, in the best cases
and for high $\tan\beta$ values where--as will be demonstrated below--%
$B^t$ can become as much as $\sim 10^{-4}$,  
is typically one-order of magnitude smaller than that of the 
diagrams in Fig.~\ref{figure1}. We will therefore assume for simplicity 
that $\lambda^{\prime}_{ \ell 3 k}=0$, thus neglecting the contribution from 
the pure $\rp$ diagram for both $\Gamma (t \to c \tilde\nu)$ and 
$\Gamma (t \to c \anb)$.

The free parameters of low energy SUSY relevant for the decays 
in question are:
$m_{\tilde\nu}$, $m_{{\tilde d}_{R}}$, $m_{{\tilde d}_{L}}$,
$m_{{\tilde \ell}_{L}}$, $m_{H^+},~m_{{\tilde\chi}_m}$, $A^{\prime}$ 
and $\tan\beta$.
To simplify our analysis below we wish to reduce the number 
of free parameters by making some plausible simplifying assumptions on the
low energy SUSY spectrum. In particular, we find that some of the above 
parameters have very little effect on $B^t$. 
We fix the values of these parameters and vary the rest: 

\begin{enumerate}

\item We find that $\Gamma (t \to c \tilde\nu)$ is practically 
insensitive to the slepton and charged Higgs masses. We therefore set
$m_{H^+}=200$ GeV and $m_{{\tilde \ell}_{L}}=m_{\tilde\nu}$. 

\item Since a possible mass 
splitting between the left and right handed down 
squarks has no effect on our scaled branching ratio $B^t$, we set 
$m_{{\tilde d}_{R}}=m_{{\tilde d}_{L}}=M_s$ for all squark flavors. 
Therefore, since $M_s$ is our typical SUSY mass scale, it is only natural 
to set $A^{\prime}$--the 
$\rp$ trilinear soft breaking term in Eq.~\ref{rplag2}--to be 
$A^{\prime}=M_s$. 

\item The two physical chargino masses 
$m_{{\tilde\chi}_m}$ ($m=1,2$) and the mixing matrices 
$Z^\pm$ are extracted by diagonalizing the chargino mass 
matrix which depends on the low energy values of the 
Higgs mass parameter $\mu$, the gaugino 
mass ${\tilde m}_2$ and $\tan\beta$ (for more details see \cite{ourtbpaper}).
It is  however sufficient to vary 
only one of the two mass parameters 
$\mu$ and ${\tilde m}_2$ in order to investigate the dependence 
of $B^t$ on the chargino masses. We, therefore, set 
${\tilde m}_2=85$ GeV. In the traditional GUT assumption, {\it i.e.,} 
that there is 
an underlying grand unification,  
${\tilde m}_2=85$ GeV corresponds to 
a gluino mass of $\sim 300$ GeV since, in that case, 
the gaugino masses are unified at 
the GUT scale leading to the relation,  
${\tilde m}_2 / m_{\rm gluino} = \alpha/ s_W^2 \alpha_s$, at the electroweak 
scale (see {\it e.g.,} \cite{ourtbpaper}). 

\item Since we are not interested here in CP violation, we take $\mu$, 
${\tilde m}_2$ and $\lambda^{\prime}$ to be real. 

\end{enumerate}

\begin{figure}[htb]
\psfull
 \begin{center}
  \leavevmode
  \epsfig{file=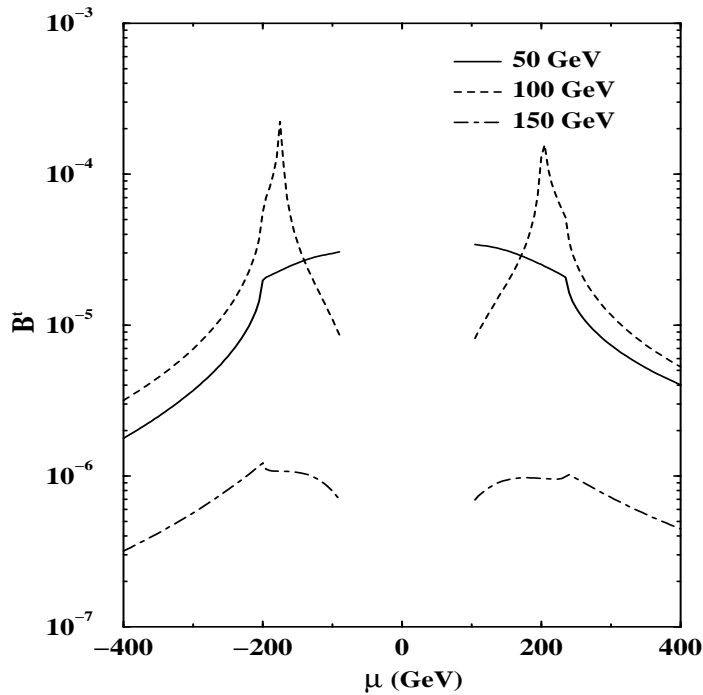,height=9.5cm,width=10cm,bbllx=0cm,bblly=2cm,bburx=20cm,bbury=25cm,angle=0}
 \end{center}
\caption{\emph{The scaled branching ratio $B^t$,
defined in Eq.~\ref{bt}, for a single neutrino flavor,
as a function  of $\mu$ for: $m_{\tilde\nu}=50$ GeV (solid line), 
$m_{\tilde\nu}=100$ GeV (dashed line) and 
$m_{\tilde\nu}=150$ GeV (dashed-dotted line). Also, $\tan\beta=35$, 
$M_s=100$ GeV, $m_{H^+}=200$ GeV, $m_{{\tilde \ell}_L}=m_{\tilde\nu}$ 
and $A^{\prime}=M_s$ are used. See also \cite{foot2} and \cite{footfig3}.}} 
\label{figure3}
\end{figure}

\begin{figure}[htb]
\psfull
 \begin{center}
  \leavevmode
  \epsfig{file=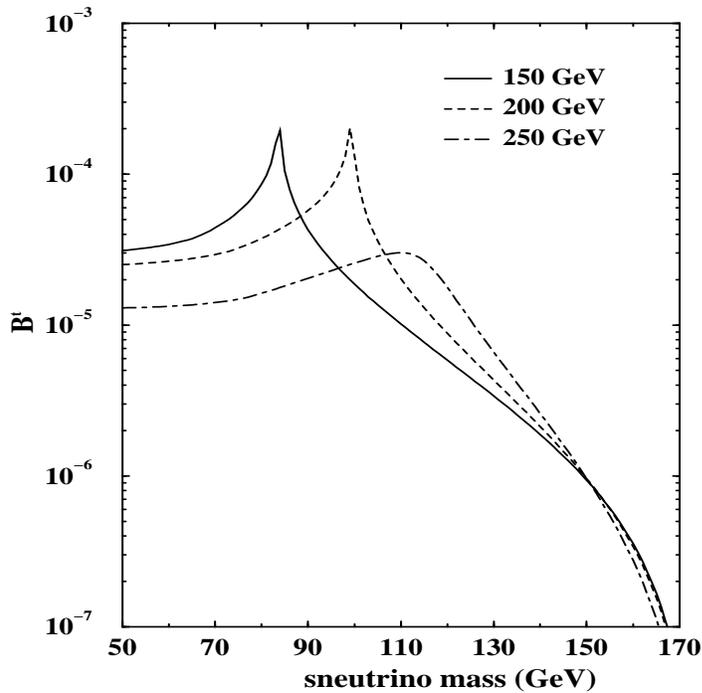,height=9.5cm,width=10cm,bbllx=0cm,bblly=2cm,bburx=20cm,bbury=25cm,angle=0}
 \end{center}
\caption{\emph{The scaled branching ratio 
$B^t$ as a function of the sneutrino mass $m_{\tilde\nu}$,   
for: $\mu = 150$ GeV (solid line), 
$\mu=200$ GeV (dashed line) and 
$\mu=250$ GeV (dashed-dotted line). 
See also caption to Fig. \ref{figure3}.}}
\label{figure4}
\end{figure}

\begin{figure}[htb]
\psfull
 \begin{center}
  \leavevmode
  \epsfig{file=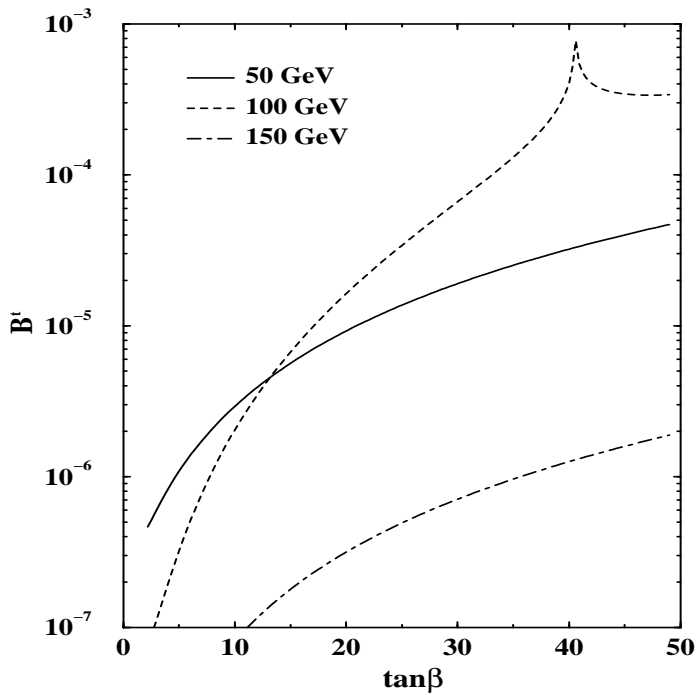,height=9.5cm,width=10cm,bbllx=0cm,bblly=2cm,bburx=20cm,bbury=25cm,angle=0}
 \end{center}
\caption{\emph{The scaled branching ratio 
$B^t$ as a function of $\tan\beta$,   
for $\mu=200$ GeV and for: $m_{\tilde\nu} = 50$ GeV (solid line), 
$m_{\tilde\nu} = 100$ GeV (dashed line) and 
$m_{\tilde\nu} = 150$ GeV (dashed-dotted line). 
The rest of the parameters 
are held fixed to their values in Figs. \ref{figure3} and \ref{figure4}. 
See also caption to Fig. \ref{figure3}.}}
\label{figure5}
\end{figure}

As it turns out, with a low $\tan\beta$, {\it i.e.,} 
$\tan\beta \lsim 10$, the branching ratio $B^t$ is typically $\lsim 10^{-6}$.
Therefore, in Figs.~\ref{figure3} and \ref{figure4}  
we focus on the high value $\tan\beta=35$. The dependence of $B^t$ on 
$\tan\beta$ is shown in Fig.~\ref{figure5}.
Also, we find that $\Gamma(t \to c \tilde\nu)$ drops as the squark 
mass scale, $M_s$, is increased,
and in what follows, we present results for $M_s=100$ GeV. 
However, it is important to note that $B^t > 10^{-6}$ is still possible 
for squark masses $\lsim 190$ GeV.   

In Fig.~\ref{figure3} we plot the scaled branching ratio $B^t$, 
as a function of $\mu$, for 
$\tan\beta=35$, $M_s=100$ GeV and for three values of the sneutrino
mass, $m_{\tilde\nu}=50,~100,~150$ GeV. Also, as stated above, here and 
throughout the rest of the paper we set $m_{H^+}=200$ GeV and 
$m_{{\tilde \ell}_{L}}=m_{\tilde\nu}$. 
We vary $\mu$ in the range 
$-400~{\rm GeV}~< \mu < 400~{\rm GeV}$ subject to 
$m_{{\tilde\chi}_m} > 50$ GeV for $m=1$ 
or 2 (see also \cite{foot2}).  
Evidently, in the range $-400~{\rm GeV}~< \mu < 400~{\rm GeV}$, 
for $m_{\tilde\nu}=50(100)$ GeV there is a $\sim 290(360)$ GeV 
range of $\mu$ in which $B^t > 10^{-5}$. Note that $B^t$ is largest 
for $m_{\tilde\nu}=100$ GeV \cite{footfig3}, reaching 
$B^t > 10^{-4}$ in the rather narrow $\mu$ mass ranges: 
$-185~{\rm GeV}~\lsim \mu \lsim -165~{\rm GeV}$ and 
$200~{\rm GeV}~\lsim \mu \lsim 210~{\rm GeV}$. 

Fig.~\ref{figure4} shows the dependence of $B^t$ on the sneutrino mass, 
for $\mu=150,~200$ and 250 GeV. Here also, $\tan\beta=35$ and $M_s=100$ GeV. 
We see that $B^t > 10^{-5}$ for $m_{\tilde\nu} \lsim 110 -125$ GeV 
depending on the value of $\mu$. Again, $B^t$ can reach above $10^{-4}$ 
in some small sneutrino mass ranges around $\sim 85$ GeV and 
$\sim 100$ GeV (see also \cite{footfig3}).

Finally, in Fig.~\ref{figure5} we show the dependence of $B^t$ on $\tan\beta$, 
for $m_{\tilde\nu}=50,~100$ and 150 GeV and for $\mu=200$ GeV. The rest 
of the parameters are fixed to the same values as in Figs.~\ref{figure3} 
and \ref{figure4}. As mentioned before, $B^t < 10^{-6}$ for small $\tan\beta$ 
values of ${\cal O}(1)$ and it increases with $\tan\beta$. It is interesting 
to note that, for $m_{\tilde\nu}=100$ GeV and with 
$\tan\beta \approx m_t/m_b \sim 40$, $B^t$ is well above $10^{-4}$ 
reaching almost $10^{-3}$ (see also \cite{footfig3}). 

To conclude this section, noting that at the LHC  
about $10^{7}-10^8$ top quarks will be produced, 
the observability of this flavor changing top quark decay will require 
at least $B^t \gsim 10^{-6}$.
Therefore, the decay $t \to c \tilde\nu$ 
may be a very good venue to determine
$\tan\beta$ in $\rp$ SUSY models, since its 
branching ratio does not reach this limit 
if $\tan\beta$ is smaller than about $10$. However, for a high 
$\tan\beta$ scenario, {\it e.g.,} with $\tan\beta \approx m_t/m_b$, 
the decay $t \to c \tilde\nu$ may have a branching ratio well above 
$10^{-5}$ in some wide ranges of the SUSY parameters, provided that 
$m_{\tilde\nu} \lsim 120$ GeV and that the squark masses are of 
${\cal O}(100)$ GeV. Although not explicitly shown above, 
we find that $B^t \gsim 10^{-6}$ 
is possible as long as the squark masses are $\lsim 190$ GeV. 
Moreover, in some small ranges 
of the SUSY parameter space, $t \to c \tilde\nu$ can have a branching ratio 
above $10^{-4}$ reaching even $10^{-3}$. 
It is interesting to note that, since the leading contribution to 
${\rm Br}(t \to c \tilde\nu)$ is independent of $m_c$ (see the term 
proportional to $|{\bar {\cal B}}^{\tilde\nu}|^2$ in Eq.~\ref{sneuwidth}),    
${\rm Br}(t \to u \tilde\nu) \simeq {\rm Br}(t \to c \tilde\nu)$ 
if $\lambda^{\prime}_{i13} \sim \lambda^{\prime}_{i23} \sim {\cal O}(1)$.

It is also useful to note that, 
with $\lambda^{\prime}_{323} \sim {\cal O}(1)$, the $\tau$-sneutrino
will decay predominantly to a pair of $b \bar s$ with 
a branching ratio of ${\cal O}(1)$ (see the next section). Therefore,    
a good way for experimentally searching for this rare 
flavor changing top decay, {\it i.e.,} $t \to c \tilde\nu$, may be to look for 
the three jets signature $t \to c b \bar s$, where the invariant $b \bar s$ 
mass reconstructs the sneutrino mass. \\

\n \underline{\bf 3. The case \boldmath{$m_{\tilde\nu} > m_t$} and 
the sneutrino decays \boldmath{$\tilde\nu \to t \bar c,~\bar t c$} and  
\boldmath{$\anb \to  t \bar c,~\bar t c$}} \\

In this section we discuss the opposite mass case, {\it i.e.,} 
$m_{\tilde\nu} > m_t$, and 
thus prospects for observing the ``reversed'' 
flavor changing sneutrino decays:

\be 
\tilde\nu_i \to t \bar c,~\bar t c ~~,~~ 
\anb_i \to t \bar c,~\bar t c \label{sneudec}~, 
\ee

\n at the LHC. 

The calculation of the widths for the sneutrino decays
in Eq.~\ref{sneudec} is straightforward using our formulas for 
$t \to c \tilde\nu_i$ and $t \to c \anb_i$ in the previous section.
For $\tilde\nu_i \to \bar t c$ $(\anb_i \to \bar t c)$ 
the amplitude is the same as the amplitude 
for $t \to c \anb_i$ $(t \to c \tilde\nu_i)$ with the top spinor 
$u_t$ replaced by the anti-top spinor $v_t$. 
We therefore have:

\bea
&& \Gamma(\tilde\nu_i \to \bar t c) = N_c 
\frac{ |{\cal C}^{\anb}|^2}{16 \pi} 
\frac{\omega(m_{{\tilde\nu}_i}^2,m_t^2,m_c^2)}{m_{{\tilde\nu}_i}^3} 
\frac{m_t^2 m_c^2}{m_W^4} 
\times \nonumber \\
&&~~~~ \left\{ \left( m_{{\tilde\nu}_i}^2 - m_t^2 - m_c^2  \right) 
\left( \frac{m_W^2}{m_t^2} |\bar {\cal A}^{\anb}|^2 +
\frac{m_W^2}{m_c^2} |\bar {\cal B}^{\anb}|^2 \right)
-4 m_W^2 \Re{\rm e} \left[\bar {\cal A}^{\anb} 
(\bar {\cal B}^{\anb})^\ast \right] \right\} \label{swidth}~,
\eea 

\n where $N_c=3$ is the number of colors, 
$\omega(m_{{\tilde\nu}_i}^2,m_t^2,m_c^2)$ is defined in 
Eq.~\ref{omega} and $\bar {\cal A}^{\anb}$,   
$\bar {\cal B}^{\anb}$ are defined in Eq.~\ref{defAB2}.

For $\anb_i \to \bar t c$ the width is given 
by Eq.~\ref{swidth} with the replacements: 
${\cal C}^{\anb} \longrightarrow 
{\cal C}^{\tilde\nu}$ and $\bar {\cal A}^{\anb}  
\longrightarrow \bar {\cal A}^{\tilde\nu}$, 
$\bar {\cal B}^{\anb}  
\longrightarrow \bar {\cal B}^{\tilde\nu}$, where 
$\bar {\cal A}^{\tilde\nu}$ and   
$\bar {\cal B}^{\tilde\nu}$ are defined in Eq.~\ref{defAB1}. 

In addition, 
if CP is a good symmetry (as is in our case since we have assumed that 
the $\lambda^{\prime}$'s, $\mu$ and ${\tilde m}_2$ are all real), then:

\bea
\Gamma(\tilde\nu_i \to t \bar c) &=& \Gamma(\anb_i \to \bar t c) 
~, \\ 
\Gamma(\anb_i \to t \bar c) &=& \Gamma(\tilde\nu_i \to \bar t c)~.
\eea  
    
\n We are now interested in the branching ratio of a sneutrino, say again the 
$\tau$-sneutrino ($i=3$) and dropping the index $i$, 
to decay to a $t \bar c$ and $\bar t c$ pairs:

\be
B^s \equiv \frac{\Gamma(\tilde\nu \to t \bar c) + 
\Gamma(\tilde\nu \to \bar t c)}{\Gamma_{tot}^{\tilde\nu}} \label{bs}~.
\ee

\n Due to CP invariance the corresponding branching ratio 
for $\anb$ equals that of $\tilde\nu$, {\it i.e.,} 
$B^s$ defined above.
It is also worth noting that, similar to the arguments given in the 
previous section, 
in our scenario with $\lambda^{\prime}_{i32} = 0$ and
 $\lambda^{\prime}_{i23} \neq 0$, 
$\Gamma(\tilde\nu \to t \bar c) > > \Gamma(\tilde\nu \to \bar t c)$ and 
$B^s \approx \Gamma(\tilde\nu \to t \bar c)/\Gamma_{tot}^{\tilde\nu}$.   

\begin{figure}[htb]
\psfull
 \begin{center}
  \leavevmode
  \epsfig{file=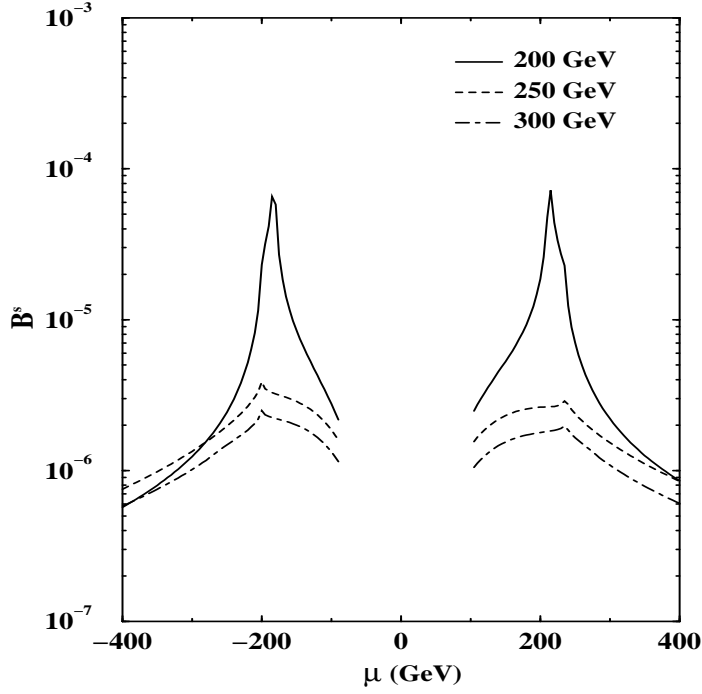,height=9.5cm,width=10cm,bbllx=0cm,bblly=2cm,bburx=20cm,bbury=25cm,angle=0}
 \end{center}
\caption{\emph{The branching ratio $B^s$, defined in Eq.~\ref{bs},
as a function 
of $\mu$, for: 
$m_{\tilde\nu}=200$ GeV (solid line), 
$m_{\tilde\nu}=250$ GeV (dashed line) and 
$m_{\tilde\nu}=300$ GeV (dashed-dotted line). 
Also, $\lambda^{\prime}_{i23}=1$ for a given sneutrino flavor $i$ and 
the rest of the parameters are set to: 
$\tan\beta=35$, 
$M_s=100$ GeV, $m_{H^+}=200$ GeV, $m_{{\tilde \ell}_L}=m_{\tilde\nu}$ 
and $A^{\prime}=M_s$. See also \cite{foot2}.}}
\label{figure6}
\end{figure}

For our purpose we define the $\tau$-sneutrino total width as:

\be
\Gamma_{tot}^{\tilde\nu} = \Gamma(\tilde\nu \to b \bar s) + 
\Gamma(\tilde\nu \to \tilde\chi \tau) + 
\Gamma(\tilde\nu \to \tilde\chi^0 \nu_\tau) \label{gamtot}~.
\ee

\n This should serve as an approximate expression for the 
$\tilde\nu$ total width since, as assumed in this analysis, 
with $m_{\tilde\nu_i} = m_{{\tilde \ell}_{Li}}$  
and when the three sneutrino flavors are degenerate, the decays
$\tilde\nu_i \to W^+ {\tilde \ell}_{Li} ,~ H^+  {\tilde \ell}_{Li}
,~ Z^0 \tilde\nu_j ,~H^0 \tilde\nu_j$ are kinematically forbidden.

\begin{figure}[htb]
\psfull
 \begin{center}
  \leavevmode
  \epsfig{file=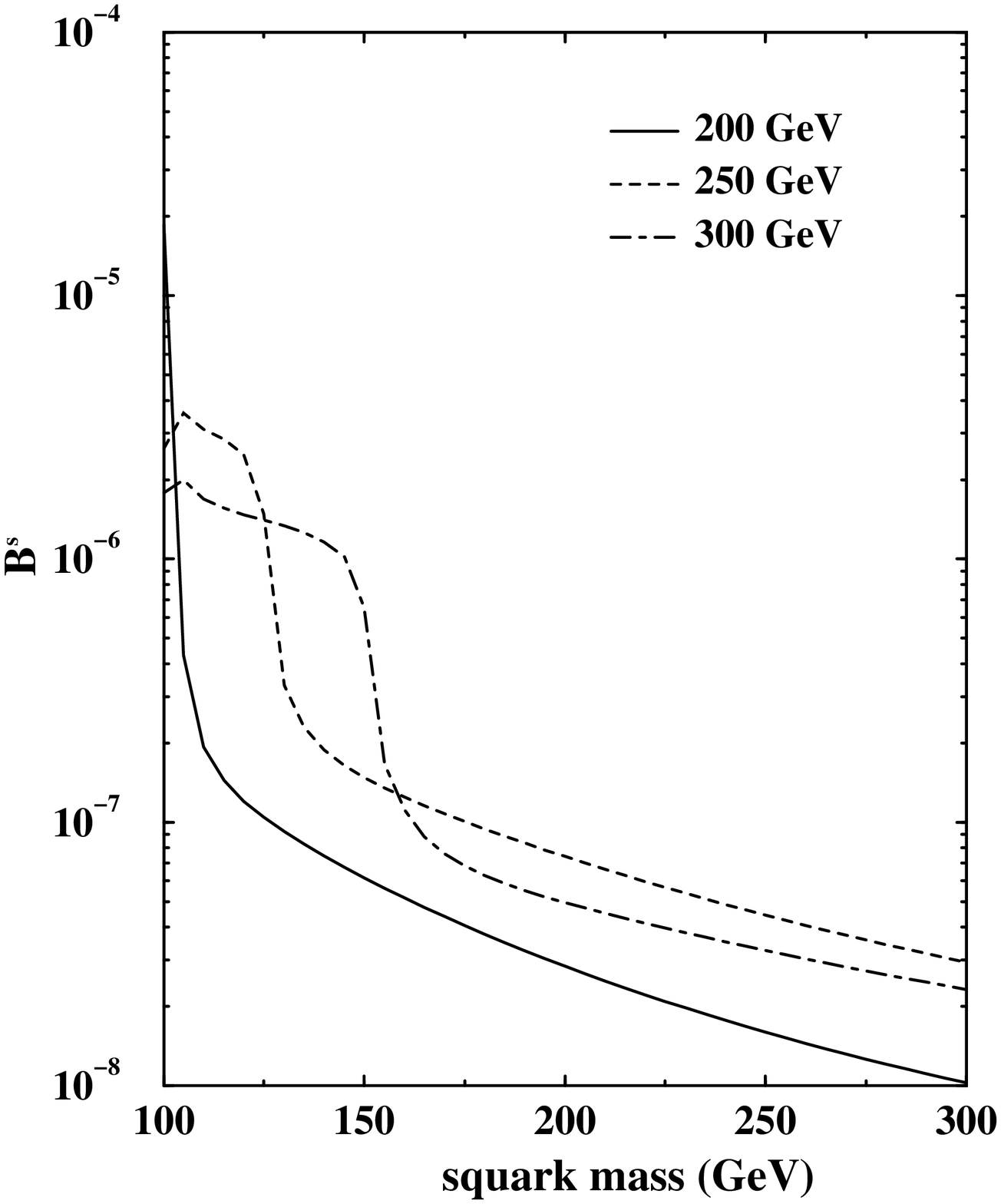,height=9.5cm,width=10cm,bbllx=0cm,bblly=2cm,bburx=20cm,bbury=25cm,angle=0}
 \end{center}
\caption{\emph{The branching ratio 
$B^s$, 
as a function 
of the squark mass $M_s$, for $\mu=200$ GeV and for: 
$m_{\tilde\nu}=200$ GeV (solid line), 
$m_{\tilde\nu}=250$ GeV (dashed line) and 
$m_{\tilde\nu}=300$ GeV (dashed-dotted line). 
See also caption to Fig. \ref{figure6}.}}
\label{figure7}
\end{figure}

The width of the $\rp$ sneutrino decay to a $b \bar s$ pair 
is given by:

\be
\Gamma(\tilde\nu \to b \bar s) = \left( \lambda^{\prime}_{323} \right)^2 
\frac{N_c}{16 \pi} m_{\tilde\nu} ~,
\ee

\n and the widths of the 
$R$-parity conserving decays in Eq.~\ref{gamtot} to a chargino ($\tilde\chi$) 
and to a neutralino ($\tilde\chi^0$) are \cite{barger}:

\be
\Gamma(\tilde\nu \to \tilde{\chi} \tau);~\Gamma(\tilde\nu \to 
\tilde{\chi}^0 \nu_\tau) = {\cal  C} \frac{g^2}{16 \pi} m_{\tilde\nu} \times 
\left(1 - m_{\tilde {\chi}}^2/m_{\tilde\nu}^2 \right)^2;~\left(1-
m_{\tilde {\chi}^0}^2/m_{\tilde\nu}^2\right)^2 \label{charneut} ~,
\ee

\n where ${\cal C} < 1$, since it is proportional to 
the square of the charginos 
or neutralinos mixing matrices \cite{barger}. Therefore, 
$\Gamma(\tilde\nu \to \tilde{\chi} \tau);~\Gamma(\tilde\nu \to 
\tilde{\chi}^0 \nu_\tau) < 10^{-2} m_{\tilde\nu}$ and, for simplicity, 
in our numerical results we conservatively ignore the additional 
phase-space factors in Eq.~\ref{charneut}, by setting  
$\Gamma(\tilde\nu \to \tilde{\chi} \tau) = \Gamma(\tilde\nu \to 
\tilde{\chi}^0 \nu_\tau) =  10^{-2} m_{\tilde\nu}$. 
Evidently, with $\lambda^{\prime}_{323} \sim {\cal O} (1)$, 
$\Gamma_{tot}^{\tilde\nu}$ is dominated by the width of the $\rp$ decay 
$\Gamma( \tilde\nu \to b \bar s) \sim 6 \times 10^{-2} m_{\tilde\nu}$.   

\begin{figure}[htb]
\psfull
 \begin{center}
  \leavevmode
  \epsfig{file=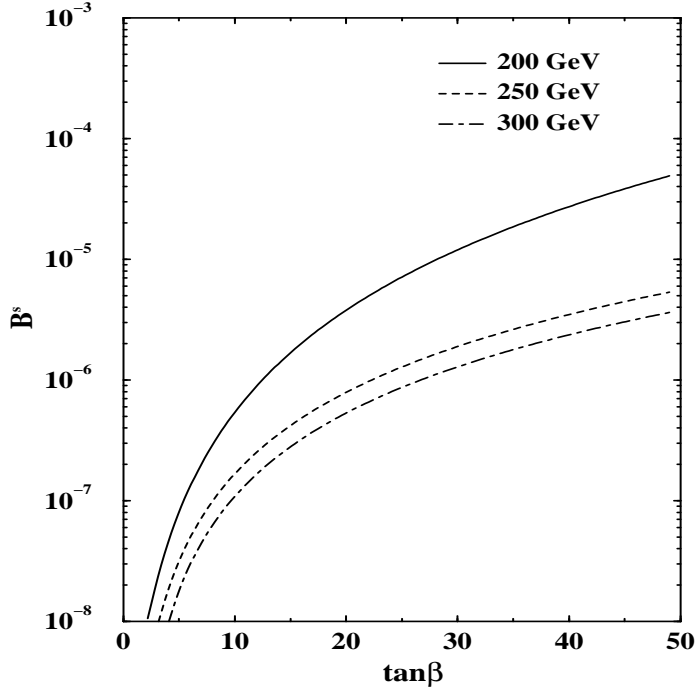,height=9.5cm,width=10cm,bbllx=0cm,bblly=2cm,bburx=20cm,bbury=25cm,angle=0}
 \end{center}
\caption{\emph{The branching ratio 
$B^s$ as a function 
of $\tan\beta$, for $\mu=200$ GeV and for: 
$m_{\tilde\nu}=200$ GeV (solid line), 
$m_{\tilde\nu}=250$ GeV (dashed line) and 
$m_{\tilde\nu}=300$ GeV (dashed-dotted line). 
See also caption to Fig. \ref{figure6}.}}
\label{figure8}
\end{figure}

Let us now discuss our numerical results 
for the branching ratio $B^s$,
defined in Eq.~\ref{bs}.
As in the previous section,
we fix $m_{H^+}=200$ GeV, $m_{{\tilde\ell}_L}=m_{\tilde\nu}$ 
($B^s$ is also found to be practically insensitive to these two parameters),
${\tilde m}_2=85$ GeV and we study the dependence of $B^s$ on 
the parameters $m_{\tilde\nu}$, $M_s$ (again setting $A^{\prime}=M_s$), 
$\mu$ and $\tan\beta$. 

In Fig.~\ref{figure6} we plot the branching ratio $B^s$ as a function 
of $\mu$ for $\lambda^{\prime}_{i23}=1$,
$\tan\beta=35$, $M_s=100$ GeV and for three values of the 
sneutrino mass, $m_{\tilde\nu}=200,~250~,300$ GeV. We see that 
for $m_{\tilde\nu}=200 - 300$ GeV, $B^s >10^{-6}$ as long as 
$|\mu| \lsim 300$ GeV. Moreover, $B^s> 10^{-5}$ in the ranges 
$-205~{\rm GeV} \lsim \mu \lsim -160~{\rm GeV}$ and    
$185~{\rm GeV} \lsim \mu \lsim 240~{\rm GeV}$.  

Although, as in the case 
of the top decay $t \to c \tilde\nu$, we find that  
in general $B^s$ decreases as the squark mass $M_s$ increases, for 
$\tilde\nu \to t \bar c,~\bar t c$ the dependence 
is not as trivial as in $t \to c \tilde\nu$. We, therefore, plot 
in Fig.~\ref{figure7} $B^s$ as a function 
of the squarks mass $M_s$, for 
$\tan\beta=35,~\mu=200$ GeV and 
$m_{\tilde\nu}=200,~250$ or 300 GeV.   
Evidently, while for $m_{\tilde\nu}=200$ GeV the branching ratio $B^s$ 
is above $10^{-7}$ only if $M_s \lsim 130$ GeV, for 
$m_{\tilde\nu}=250$ GeV and $m_{\tilde\nu}=300$ GeV 
$B^s$ can be  $> 10^{-7}$ 
for larger $M_s$ values: $M_s \lsim 180$ GeV and   
$M_s \lsim 160$ GeV, respectively.

In Fig.~\ref{figure8} we show the dependence of 
$B^s$, on $\tan\beta$, for the three sneutrino masses 
$m_{\tilde\nu}=200,~250$ and 300 GeV and with $\lambda^\prime_{i23}=1$,
$M_s=100$ GeV and $\mu=200$ GeV. 
The same behavior as in the case of $t \to c \tilde\nu$ 
is found. In particular, $B^s < 10^{-7}$ for $\tan\beta \sim {\cal O}(1)$ 
and it increases with $\tan\beta$ such that $B^s > 10^{-7}$ for 
$\tan\beta \gsim 10$. For $m_{\tilde\nu}=200$ GeV and 
$\tan\beta \gsim 30$, $B^s > 10^{-5}$.     

Note that, as in the case of the top decays and for the same 
reasons, the results above apply also to 
${\rm Br}(\tilde\nu \to t \bar u)$ if instead
$\lambda^{\prime}_{i13} \sim {\cal O}(1)$. \\

\n \underline{\bf 3. Summary and conclusions} \\

We have calculated the branching ratios for 
the flavor changing top quark decays $t \to c \tilde\nu,~c \anb$ 
when $m_{\tilde\nu} < m_t$, or for the corresponding reversed sneutrino 
decays $\tilde\nu \to t \bar c,~\bar t c$ and    
$\anb \to t \bar c,~\bar t c$ when $m_{\tilde\nu} > m_t$, 
in the MSSM with $R$-parity violation. In this model, these decays 
can occur at the one-loop level and they depend predominantly 
on the squark masses, the Higgs mass parameter $\mu$, $\tan\beta$ and 
for a sneutrino flavor $i=e,~\mu$ or $\tau$, 
on the $\rp$ couplings $\lambda^{\prime}_{i23}$. 

We have considered the values $\lambda^{\prime}_{i23} \sim {\cal O}(1)$, 
which are not ruled out if one 
allows more than one $\rp$ coupling to be non-zero, and  
showed that 
these rare decays are sensitive probes of the 
parameter $\tan\beta$, since their branching ratios become 
experimentally accessible  
({\it i.e.,} at the LHC), since typically, 
${\rm Br}(t \to c \tilde\nu) \gsim 10^{-6}$ and 
${\rm Br}(\tilde\nu \to t \bar c) \gsim 10^{-7}$ only for 
$\tan\beta \gsim 10$. 

For the top decays in the case $m_{\tilde\nu} < m_t$, we found that 
 $\Gamma(t \to c \tilde\nu) >> \Gamma(t \to c \anb)$ and 
that ${\rm Br}(t \to c \tilde\nu)$ can be well within the reach of the LHC 
with $10^7 -10^8$ $t \bar t$ pairs produced. 
In particular, it was shown that 
${\rm Br}(t \to c \tilde\nu) > 10^{-5}$ for $\tan\beta \gsim 30$ 
in a $\gsim 300$ GeV range of the Higgs mass parameter $\mu$ 
as long as the squark masses are not much larger than 100 GeV. 
In some cases, with $\tan\beta \sim m_t/m_b \sim 40$, we found that  
${\rm Br}(t \to c \tilde\nu) > 10^{-4}$ reaching almost $10^{-3}$. 

For the reversed sneutrino decays in the case $m_{\tilde\nu} > m_t$, 
we found that similar to the above mentioned $t$ decays, 
$\Gamma(\tilde\nu \to t \bar c) >> \Gamma(\tilde\nu \to \bar t c)$. 
Furthermore, ${\rm Br}(\tilde\nu \to t \bar c) \gsim 10^{-6}$ 
for $ |\mu| \lsim 300$ GeV, $\tan\beta \gsim 30$, again,  
as long as the squarks have masses around 100 GeV. 
Here also the branching ratio can be more than an order of magnitude larger,
{\it e.g.,} for $\tan\beta \sim m_t/m_b \sim 40$ and squark masses around 
100 GeV,   
${\rm Br}(\tilde\nu \to t \bar c) \sim 10^{-5} - 10^{-4}$ is possible. 
As mentioned in the introduction, 
the LHC will be able to produce $10^8 -10^9$ sneutrinos with a mass 
of $200 - 300$ GeV, if indeed 
some of the $\rp$ couplings of the $\lambda^{\prime}$ type 
are saturated to be of ${\cal O}(1)$. Therefore, if 
$t \to c \tilde\nu$ is not detected at the LHC, then  
our results above indicate that it may still be useful 
to search for the reversed sneutrino decay 
$\tilde\nu \to t \bar c$ via the reaction 
$pp \to \tilde\nu +X \to t \bar c +X$. 

Finally, changing the charm quark to an up 
quark in these decays has a negligible effect on the branching ratios 
since, by assumption, the couplings are flavor independent. 
Therefore, if in addition to 
$\lambda^{\prime}_{i23} \sim {\cal O}(1)$ also
$\lambda^{\prime}_{i13} \sim {\cal O}(1)$, then 
the same results are obtained for the decays 
$t \to u \tilde\nu,~u \anb$ and $\tilde\nu \to t \bar u,~ \bar t u$.

\bigskip
\bigskip

We acknowledge partial support from the US Israel BSF (G.E. and A.S.) 
and from the US DOE contract numbers DE-AC02-76CH00016(BNL) (A.S.), 
DE-FG03-94ER40837(UCR) (S.B.). 
G.E. also thanks the
VPR Fund at the Technion for partial support.

%

%
%
%
%






\begin{thebibliography}{99}
\bibitem{foot1} We note 
however, that in order to prevent fast 
proton decay one has to assume that either 
baryon number or lepton number is conserved 
or violated to an extremely small degree \cite{rpreview}.
\bibitem{footfig1} Note that diagrams with off-diagonal self-energy 
transitions followed by sneutrino emission off the top or charm legs 
are absent 
at one-loop, since gauge invariance forbids $\rp$ sneutrino couplings 
to a pair of up type quarks at tree-level.
\bibitem{foot2} Note that, since we limit the chargino mass $m_{{\tilde\chi}_m} < 50$ GeV, in those cases where the 
values of $\mu$ together with ${\tilde m}_2=85$ GeV and $\tan\beta=35$ 
give $m_{{\tilde\chi}_m} < 50$ GeV for $m=1$ 
or 2, the lines in Fig.~\ref{figure3} end rather abruptly.
\end{thebibliography}
\end{document}